\documentclass[aps,prl,superscriptaddress,showkeys,showpacs,twocolumn,a4paper]{revtex4-1}
\usepackage{ulem}
\usepackage{color}
\usepackage{graphicx}
\usepackage{appendix}
\usepackage{epsfig}
\usepackage[colorlinks=true,linkcolor=blue]{hyperref}
\usepackage{epstopdf}
\usepackage{amsmath}
\usepackage{subfig}
\usepackage{comment}
\usepackage[dvipsnames]{xcolor}
\newcommand{\be}{\begin{equation}}
\newcommand{\ee}{\end{equation}}
\newcommand{\ba}{\begin{eqnarray}}
\newcommand{\ea}{\end{eqnarray}}

\begin{document}
\title{Electromagnetic response of hot QCD medium in the presence of background time-varying fields}
\author{Gowthama K K}
\email{k$_$gowthama@iitgn.ac.in}
\affiliation{Indian Institute of Technology Gandhinagar, Gandhinagar-382355, Gujarat, India}

\author{Manu Kurian}
\email{manu.kurian@iitgn.ac.in}
\affiliation{Indian Institute of Technology Gandhinagar, Gandhinagar-382355, Gujarat, India}

\author{Vinod Chandra}
\email{vchandra@iitgn.ac.in}
\affiliation{Indian Institute of Technology Gandhinagar, Gandhinagar-382355, Gujarat, India}

\begin{abstract}
The response of the hot QCD medium in the presence of external time dependent electromagnetic fields has been studied within the relativistic Boltzmann transport theory. The impact of the time dependence of the electromagnetic fields and collisional aspects of the medium to the induced electric and Hall current densities has been explored. The non-equilibrium momentum distribution of degrees of freedom has been obtained in the presence of space-time varying electromagnetic fields. Further, the analysis has been extended to an anisotropic QCD medium while incorporating the in-medium interaction effects. It is observed that the electric charge transport is sensitive to the inhomogeneity of the fields and the momentum anisotropy of the QCD medium.
\end{abstract}
\maketitle

 \section{Introduction}
Experiments at Relativistic Heavy Ion Collider (RHIC) and Large Hadron Collider (LHC) have provided evidence for the existence of strongly interacting matter-Quark Gluon plasma (QGP)~\cite{Adams:2005dq}. Transport coefficients of the hot QCD matter serve as the input parameters for the hydrodynamical description of the evolution of the created medium and act as a key ingredient to explore critical properties of the medium~\cite{Jaiswal:2020hvk}. Recent observations at the RHIC and LHC have indicated the presence of a strong magnetic field in the early stages of heavy-ion collision~\cite{Acharya:2019ijj,Adam:2019wnk}. 
Even though there is no definitive description of the evolution of the generated magnetic field, various models predict that the magnetic field may persist for much longer in the QGP due to the back reaction in the medium~\cite{Tuchin:2013apa,McLerran:2013hla, Stewart:2021mjz,Tuchin:2019gkg,Yan:2021zjc}. This persisting magnetic field may influence the behavior of the hot QCD matter and hence is pertinent to study the various properties of the QCD medium in the presence of magnetic fields.  

The electric charge transport in the hot deconfined quark matter has gained much attention due to the generated electromagnetic fields in the collision experiments. The medium response to the external electromagnetic fields can be studied in terms of the induced electric and Hall current densities and associated conductivities. The generated electric field in asymmetric heavy-ion collisions has a preferred direction, and the conductivities may strongly depend on the charge asymmetric flow~\cite{Hirono:2012rt}. The study of electromagnetic responses on the QCD medium is also relevant in the context of Chiral Magnetic Effect~\cite{PhysRevD.78.074033} and emission rate of soft photons~\cite{Yin:2013kya} in the heavy-ion collisions. Electric charge transport has been studied in the presence of constant electric and magnetic fields in the regimes where the strengths of the fields are weak~\cite{Feng:2017tsh,Das:2019wjg,Thakur:2019bnf,Dey:2019axu,Kurian:2020qjr,Dash:2020vxk} and in the strong field limit~\cite{Hattori:2016lqx,Fukushima:2017lvb,Kurian:2017yxj,Ghosh:2019ubc}. Attributing to the fact that the generated electromagnetic fields are varying with space and time~\cite{Deng:2012pc,Tuchin:2013apa,Hongo:2013cqa,Li:2016tel,Huang:2015oca}, it is important to investigate the QCD medium response to the inhomogeneous fields. This sets the motivation of the current study.

Electrical conductivity of the hot QCD medium has been investigated in several studies within transport theory~\cite{Greif:2014oia,Cassing:2013iz,Puglisi:2014sha}, Kubo formalism~\cite{Kubo:1957mj} and lattice QCD estimations~\cite{Astrakhantsev:2019zkr,Amato:2013naa}. Recently, we have estimated the additional component to the current density due to the time dependence of the external electric field~\cite{Gowthama:2020ghl}. The focus of the current investigation is to extend the analysis to develop a general framework to study the response of the QCD medium to a time-varying electric and magnetic field. To that end, we have obtained the general form of the near-equilibrium quark degrees of freedom in the presence of inhomogeneous fields within the transport theory. We have also explored the electric charge transport of an anisotropic weakly magnetized QCD medium. The in-medium interactions have been incorporated in the analysis by adopting a recently developed effective fugacity quasiparticle model (EQPM)~\cite{Chandra:2011en}. We have analyzed the impact of time dependence of the fields and momentum anisotropy of the medium to the electric charge transport process along with the effects of mean-field corrections to the respective current densities.

This may perhaps be the first attempt where the physics of inhomogeneity of the external electromagnetic fields has been incorporated in the analysis of electric charge transport in the context of QCD medium.  Our estimations are consistent with the studies so far, as we have correctly reproduced the earlier results within the general formalism with the choice of appropriate electromagnetic fields.

\section{Electric charge transport in QCD medium with time-varying fields}
The response of the QCD medium to the external electromagnetic fields can be studied with induced current densities and the associated conductivities. The general form of the induced vector current in the QCD medium with a non-vanishing quark chemical potential $\mu$ in terms of quark and antiquark momentum distribution function $f_k=f^0_k+\delta f_k$ (the subscript $k$ denotes the particle species) is as follows,
\begin{align}\label{1}
j^i&=2N_c\sum_f \int d Pv^i\Big(q_q f_q-q_{\bar{q}}f_{\bar{q}}\Big),  
 \end{align}
where $v_i$ is the component of velocity and $d P = \frac{d^3 \textbf{p}}{(2\pi)^3}$.
The flavor summation (over the up, down, and strange quarks) arises from the degeneracy factor $2N_c\sum_f$ of the quarks/antiquarks with $N_c$ number of colors.
The magnetic field dependence to the current densities are entering through the non-equilibrium part of the distribution function $\delta f_k$ via cyclotron motion and has an impact on the associated conductivities. We proceed further to find the near-equilibrium distribution function by solving the transport equation within the relaxation time approximation. The relativistic transport equation that describes the dynamics of the distribution function has the following form,
\begin{align}\label{2}
{p}^{\mu}\,\partial_{\mu}f_k(x,{p})+q_{f_k}F^{\mu\nu}{p}_{  \nu}\partial^{(p)}_{\mu} f_k=-(u.p)\frac{\delta f_k}{\tau_R},
\end{align}
where $q_{f_k}$ is the electric charge of the quark/antiquark with flavor $f$, $u^{\mu}$ is the fluid velocity, $\tau_R$ is the relaxation time, taken from~\cite{Hosoya:1983xm}, and $F^{\mu\nu}$ is the electromagnetic strength tensor. We solve the relativistic Boltzmann equation by assuming the following ansatz for $\delta f_k$ due to the inhomogeneous electromagnetic fields,
\begin{equation}\label{3}
\delta f_k=({\bf{p}}.{\bf \Xi} ) \frac{\partial f^0_k}{\partial \epsilon}, 
\end{equation} 
where the vector ${\bf{\Xi}}$ is related to the strength of the electromagnetic field and its first-order (leading order) spacetime derivatives, with the following form,
\begin{align}\label{4}
\mathbf{\Xi} =& \alpha_1\textbf{E}+ \alpha_2\dot{\textbf{E}}+ \alpha_3(\textbf{E}\times \textbf{B})+ \alpha_4(\dot{\textbf{E}}\times \textbf{B})+ \alpha_5(\textbf{E}\times \dot{\textbf{B}})\nonumber\\&+\alpha_6 ({\pmb \nabla} \times \textbf{E}) +\alpha_7 \textbf{B}+\alpha_8 \dot{\textbf{B}}+\alpha_9 ({\pmb \nabla} \times \textbf{B}).
\end{align} 
Here, $\alpha_{i}$ ($i=(1, 2,.., 9)$) are the unknown functions that relate to the respective electric charge transport coefficients and can be obtained by the microscopic description of the QCD medium. The present focus is on the case with chiral chemical potential is zero and hence $\alpha_{i}=0$ for $i=(6,7,8)$ considering the parity symmetry in the analysis~\cite{Satow:2014lia}. 
Employing Eq.~(\ref{3}) and Eq.~(\ref{4}) in Eq.~(\ref{2}) we have,
\begin{widetext}
\begin{align}\label{6}
& \epsilon {\bf v}.\Big[\alpha_1 \dot{{\bf E}}+\dot{\alpha_1 }{\bf E}+\alpha_2 \ddot{{\bf E}}+\dot{\alpha_2 }\dot{{\bf E}}+\alpha_3 (\dot{{\bf E}}\times {\bf B}) +\alpha_3 ({\bf E}\times \dot{{\bf B}})+\dot{\alpha_3 }({\bf E}\times {\bf B})+\alpha_4 (\dot{{\bf E}}\times \dot{{\bf B}})+\alpha_4 (\ddot{{\bf E}} \times {\bf B}) +\dot{\alpha_4 }(\dot{{\bf E}} \times {\bf B}) \nonumber\\
&+\alpha_5 (\dot{{\bf E}}\times \dot{{\bf B}})+\alpha_5 ({\bf E}\times \ddot{{\bf B}})+\dot{\alpha_5 }({\bf E}\times \dot{{\bf B}})+\alpha_9 (\pmb{\nabla} \times {\bf \dot{B}})+\dot{\alpha_9}(\pmb{\nabla} \times {\bf B})\Big]
 +q_{f_k}{\bf v}.{\bf E}-\alpha_1 q_{f_k}{{\bf v}}.({\bf E}\times {\bf B})-\alpha_2 q_{f_k}{\bf v}.(\dot{{\bf E}}\times {\bf B})\nonumber\\
& +\alpha_3 q_{f_k}({\bf v}.{\bf E})(B^2)-\alpha_3 q_{f_k}({\bf v}.{\bf B})({\bf B}.{\bf E})+\alpha_4 q_{f_k}({\bf v}.\dot{{\bf E}})(B^2)-\alpha_4 q_{f_k}({\bf v}.{\bf B})({\bf B}.\dot{{\bf E}})+\alpha_5 q_{f_k}({\bf v}.{\bf E})(\dot{{\bf B}}.{\bf B}) -\alpha_5 q_{f_k}(\dot{{\bf B}}.{\bf v})({\bf E}.{\bf B})\nonumber\\
&-\alpha_9 q_{f_k}({\bf B.v})(\pmb{\nabla}.{\bf B})=-\frac{\epsilon}{\tau_R}\Big[\alpha_1 {\bf v}.{\bf E}+\alpha_2 {\bf v}.\dot{{\bf E}}+\alpha_3 {\bf v}.({\bf E}\times {\bf B})+\alpha_4 {\bf v}.(\dot{{\bf E}}\times {\bf B})+\alpha_5 {\bf v}.({\bf E}\times \dot{{\bf B}})+\alpha_9 {\bf v}.(\pmb{\nabla} \times {\bf B})\Big],
\end{align}
\end{widetext}
with $B=|{\bf B}|$. We consider the terms with first-order derivatives of the fields and neglect higher-order derivative terms as the electromagnetic fields vary slowly in space and time to incorporate the collisional aspects of the QCD medium. Hence, the terms with $\dot{\alpha_2}, \dot{\alpha_4}, \dot{\alpha_5}, \dot{\alpha_9}$ are neglected in the current analysis. Incorporating this approximation and comparing the coefficients of terms with the same  tensorial structure in both sides of Eq.~(\ref{6}), we obtain the coupled differential equations as follows,
\begin{align}\label{9}
  & \dot{\alpha_1} =  -\bigg[\frac{1}{\tau_R}\alpha_1 +(\frac{q_{fk}B^2}{\epsilon} -\frac{\tau_R q_{fk}B\dot{B}}{\epsilon})\alpha_3 +\frac{q_{fk}}{\epsilon}\bigg],\nonumber \\  
  & \dot{\alpha_3} = -\frac{1}{\tau_R}\alpha_3 +\frac{q_{fk}}{\epsilon}\alpha_1,
\end{align}
along with the coupled equations,
\begin{align}\label{10}
&\alpha_2=-\tau_R\Big[ \alpha_1 +\frac{q_{f_k} \alpha_4 B^2}{\epsilon}\Big],
 &&\alpha_5 = -\tau_R \alpha_3,\nonumber \\
     &{\alpha_4}=-{\tau_R}\Big[\alpha_3-\frac{\alpha_2 q_{f_k}}{\epsilon}\Big].
\end{align}
with $\dot{B}=|{\bf \dot{B}}|$. The coupled differential equations can be described in terms of matrix equation as follows, 
\begin{equation}\label{12}
 \frac{d X}{d t}= AX +G,   
\end{equation}
where the matrices can be defined as follows,
\begin{align}
 X=\begin{pmatrix}
\alpha_1\\\alpha_3
\end{pmatrix},  
&& A=\begin{pmatrix}
 -\frac{1}{\tau_R} & -\frac{q_{f_k} F^2}{\epsilon}\\
\frac{q_{f_k}}{\epsilon} &  -\frac{1}{\tau_R}, 
\end{pmatrix},
&&& G=\begin{pmatrix}
\frac{-q_{f_k}}{\epsilon}\\ 0
\end{pmatrix},\nonumber
\end{align}
with $F = \sqrt{B(B-\tau_R \dot{B})}$.
We solve Eq.~(\ref{12}) by finding the solution of the homogeneous equation $\frac{dX}{dt}=A X$ by diagonalizing the matrix $A$. 
Then, we find the solution to the non-homogeneous part  of Eq.~(\ref{12}) by replacing the constants of integration with time dependent functions $k_1(t)$ and $k_2(t)$. Using the method of variation of constants, we obtain the particular solution as follows,
\begin{align}\label{16}
&\alpha_1 =k_1 iFe^{\eta_1} -k_2 iFe^{\eta_2},
&&\alpha_3=k_1 e^{\eta_1} +k_2 e^{\eta_2}. 
\end{align}
The functions $k_1(t)$ and $k_2(t)$ can be defined as $k_1 =\frac{iq_{f_k}}{2\epsilon} I_1$ and  $k_2 =-\frac{iq_{f_k}}{2\epsilon} I_2$ where $\eta_j$ and $I_j$, $(j=1, 2)$ take the following forms, 
\begin{align}\label{21}
    &\eta_j = -\frac{t}{\tau_R} +a_j\frac{q_{f_k} i}{\epsilon}\int F dt,
    &&I_j = \int \frac{ e^{-\eta_j}}{F},
\end{align}
with $a_1=1$ and $a_2=-1$. Substituting Eq.~(\ref{21}) in Eq.~(\ref{16}) and employing Eq.~(\ref{10}), we obtain the master equation for $\alpha_i$ as follows,
\begin{align}\label{20}
 &\alpha_1 =-\frac{\Tilde{\Omega}_k}{2}(I_1 e^{\eta_1} +I_2 e^{\eta_2} ),\\
 &\alpha_3 =\frac{q_{fk}i}{2\epsilon}(I_1 e^{\eta_1} -I_2 e^{\eta_2} ),\label{20.2}\\
 &\alpha_5 =-\frac{\tau_R q_{fk}i}{2\epsilon}(I_1 e^{\eta_1} -I_2 e^{\eta_2}  ),\label{20.4}\\
 &\alpha_2 =\frac{(\frac{\Tilde{\Omega}_k\tau_R}{2} +\frac{i\Omega^2_k\tau^2_R}{2})I_1 e^{\eta_1}+(\frac{\Tilde{\Omega}_k\tau_R}{2}-\frac{i\Omega^2_k\tau^2_R}{2}) I_2 e^{\eta_2}}{1+\Omega^2_k\tau^2_R}\label{20.1}\\
 &\alpha_4 =\frac{-(\frac{\Tilde{\Omega}^2_k\tau^2_R}{2F} +\frac{iq_{fk}\tau_R}{2\epsilon})I_1 e^{\eta_1}-(\frac{\Tilde{\Omega}^2_k\tau^2_R}{2F}-\frac{iq_{fk}\tau_R}{2\epsilon}) I_2 e^{\eta_2}}{1+\Omega^2_k\tau^2_R},\label{20.3}
\end{align}
with $\Omega_k=\frac{q_{f\, k}B}{\epsilon}$ represents the cyclotron frequency at finite $B$ and $\Tilde{\Omega}_k=\frac{q_{f\,k}F}{\epsilon}$ for the time-varying magnetic field. Now, we solve the master equation of $\alpha_i$ for various choices of the electromagnetic fields \textbf{E} and \textbf{B}.
\subsubsection*{Case I: Constant electric and magnetic fields}
For the case of constant electric and magnetic field, the terms with $\alpha_1$ and $\alpha_3$ will give non-zero contributions to the current density as the terms associated with space-time derivatives of the fields vanish as described in Eq.~(\ref{3}). Solving $\alpha_1$ and $\alpha_3$ from Eq.~(\ref{20}) and Eq.~(\ref{20.2}) for constant electromagnetic fields, we obtain 
\begin{align}\label{25}
 &\alpha_1 = \frac{-\epsilon q_{f_k}}{\tau_R [(\frac{\epsilon}{\tau_R})^2 +(q_{f_k} B)^2]},
 &&\alpha_3 =\frac{-q_{f_k}^2 }{[(\frac{\epsilon}{\tau_R})^2 +(q_{f_k} B)^2]}.
\end{align}
Employing Eq.~(\ref{3}) in Eq.~(\ref{1}), we obtain  $j^i=\sigma_{e} \delta^{ij} E_j +\sigma_H \epsilon^{ij}E_j$ with $\sigma_{e}$ and $\sigma_H$ denote the electrical and Hall conductivites, respectively. The results obtained are in agreement with the observations of Ref.~\cite{Feng:2017tsh}. It is important to emphasize that the Hall current vanishes at $\mu=0$ case. In the limit of vanishing magnetic field, we obtain the electric current as,
\begin{align}
{\bf j}=\frac{{\bf E}}{3} 2N_c\sum_k \sum_f (q_{f_k})^2 \int dP \tau_R\frac{p^2}{\epsilon^2}(-\frac{\partial f^0_k}{\partial \epsilon}),
\end{align}
which agrees with the results of~\cite{Puglisi:2014sha}. However, in the presence of a strong uniform magnetic field, the longitudinal electrical conductivity has the dominant role in the electric charge transport and $\sigma_H\sim 0$ as the transverse transport to the magnetic field is negligible due to the $1+1-$D Landau dynamics of quarks and antiquarks.
\subsubsection*{Case II: Response of weakly magnetized medium to time dependent electric field}
We consider the electric charge transport process in the presence of a time-varying electric field in a weakly magnetized QCD medium.
Hence, the term with $\textbf{E} \times \dot{\textbf{B}}$ will not contribute to the current density. In the case where $\dot{\textbf{B}}=0$, we have $F = B$ and Eq.~(\ref{21}) reduces to,
\begin{align}\label{27}
    &\eta_j = \big(-\frac{1}{\tau_R} +a_j\Omega_ki\big)t,
    &&I_j = \frac{-e^{ -\big(-\frac{1}{\tau_R} +a_j\Omega_ki\big)t }}{B(-\frac{1}{\tau_R} +a_j\Omega_ki)}.
\end{align}
Substituting Eq.~(\ref{27}) in master equations Eqs.~(\ref{20})-(\ref{20.3}), we obtain non-zero contribution for the current density from the terms with $\alpha_i,(i=1,2,3,4)$.
The coefficient $\alpha_1$ and $\alpha_3$ that relate to the electrical and Hall conductivities are defined in Eq.~(\ref{25}). The additional parameters that arise due to the time dependence of the external electric field take the following forms,
\begin{align}\label{28}
 &\alpha_2 =\frac{q_{f_k} \epsilon [(\frac{\epsilon}{\tau_R})^2 -(q_{f_k} B)^2]}{[(\frac{\epsilon}{\tau_R})^2 +(q_{f_k} B)^2]^2},
 &&\alpha_4 =\frac{2q_{f_k}^2 \epsilon^2 }{\tau_R [(\frac{\epsilon}{\tau_R})^2 +(q_{f_k} B)^2]^2}.
\end{align}
By employing Eq.~(\ref{3}) in Eq.~(\ref{1}), we can obtain an additional component to the Ohmic current due to the time dependence of the external electric field and can be quantified in terms of $\alpha_2$. Note that the additional component is higher-order in $\tau_R$ in comparison to the Ohmic current density. Similarly, $\alpha_4$ is related to the additional component to the Hall current in the presence of time-varying electric field. The results obtained are consistent with that of Ref.~\cite{Gowthama:2020ghl}. 
\subsubsection*{Case III: Response to time-varying electromagnetic field}
In the case where both electric and magnetic fields are dependent on time, the contributions to the current density from all the terms associated with $\alpha_i, (i=1,2,3,4,5)$ need to be considered systematically. We begin with evaluating the integrals described in Eq.~(\ref{21}) by choosing a particular time dependence of the external magnetic field. We consider the magnetic field with the form $\textbf{B} = B_0 e^{-\frac{t}{\tau_B}}\hat{\bf z}$, where $B_0$ is its amplitude and $\tau_B$ is the decay time parameter~\cite{Hongo:2013cqa, Satow:2014lia} such that $F = B\sqrt{1+\frac{\tau_R}{\tau_B}}$, in the current analysis. Within the limit that the cyclotron frequency $\Omega_k$ is approximately equal to the decay frequency ($\tau_B^{-1}$) of the magnetic field, Eq.~(\ref{21}) reduces to the following form,
\begin{align}\label{30}
    &\eta_j = -\frac{t}{\tau_R} +a_j\Bigg(i\frac{\sqrt{1+\frac{\tau_R}{\tau_B}}}{\tau_B}t\Bigg),
\end{align}
\begin{align}
    &I_j = \frac{1}{\sqrt{1+\frac{\tau_R}{\tau_B}} B_0} \frac{e^{\Big(\frac{1}{\tau_R} +\frac{1}{\tau_B}-a_ji\frac{\sqrt{1+\frac{\tau_R}{\tau_B}}}{\tau_B} \Big)t}}{\Big( \frac{1}{\tau_R} +\frac{1}{\tau_B}-a_ji\frac{\sqrt{1+\frac{\tau_R}{\tau_B}}}{\tau_B} \Big)}.\label{30.2}
\end{align}
Further, we proceed with the estimation of all $\alpha_i$ coefficients by substituting Eq.~(\ref{30}) and Eq.~(\ref{30.2}) in Eq.~(\ref{20})-(\ref{20.3}). Incorporating the non-zero contributions associated with $\alpha_i, (i=1,2,..5)$ in Eq.~(\ref{4}), we obtain five components of the induced current ${\bf j}=j_e{\bf\hat{e}}+j_H({\bf\hat{e}}\times{\bf\hat{b}})$ as follows,
\begin{align}
    &j_e = j_e^{(0)}+j_e^{(1)}, &&j_H = j_H^{(0)}+j_H^{(1)}+j_H^{(2)}, 
\end{align}
where $j_e$ corresponds to the electric current in the direction of the electric field ${\bf\hat{e}}$ and $j_H$ is the electrical current in the direction perpendicular to both electric and magnetic fields $({\bf\hat{e}}\times{\bf\hat{b}})$ with
\begin{align}\label{31}
    &j_e^{(0)} =  \frac{2E}{3} N_c \sum_k \sum_f (q_{fk})^2 \int dP\frac{ p^2}{\epsilon^2}(-\frac{\partial f_k^0}{\partial \epsilon}) {M}_1,\\ 
    &j_e^{(1)} =\frac{2\dot{E}}{3} N_c \sum_k \sum_f (q_{fk})^2\int dP \frac{ p^2}{\epsilon^2}\frac{\partial f_k^0}{\partial \epsilon}M_2,\label{31.1} \\
     &j_H^{(0)} = \frac{2E}{3} N_c \sum_k \sum_f (q_{fk})^3\int dP \frac{ p^2}{\epsilon^3}(-\frac{\partial f_k^0}{\partial \epsilon}) M,\label{31.2}\\ 
     &j_H^{(1)} =\frac{2\dot{E}}{3} N_c \sum_k \sum_f (q_{fk})^3\int dP \frac{ p^2}{\epsilon^3}\frac{\partial f_k^0}{\partial \epsilon}M_3,\label{31.3}\\ 
     &j_H^{(2)} = \frac{2E}{3\tau_B} N_c \sum_k \sum_f (q_{fk})^3\int dP \frac{ p^2}{\epsilon^3}(-\frac{\partial f_k^0}{\partial \epsilon}) \tau_R M,\label{31.4}
\end{align}
where $E=|{\bf E}|$, $\dot{E}=|{\bf \dot{E}}|$ and $M_j (j=1, 2, 3)$ functions can be defined as $M_1 =\big(\frac{1}{\tau_R} +\frac{1}{\tau_B}\big) M$, $M_2={\big(\tau_R M_1  -\frac{\tau_R^2 }{\tau_B^2 } M\big)}/\big({1+(\frac{\tau_R}{\tau_B})^2}\big)$ and $M_3=\big({\tau_R M + \tau_R^2M_1 \big)}/\big({1+(\frac{\tau_R}{\tau_B})^2}\big)$ with, 
\begin{align}
 M = \Bigg[{ \frac{1}{ \tau_R} +\frac{1}{\tau_B}+\frac{\sqrt{1+\frac{\tau_R}{\tau_B}}}{\tau_B} }\Bigg]^{-1}.
\end{align}
In Eqs.~(\ref{31})-(\ref{31.4}), $j_e^{(0)}$ denotes the leading-order Ohmic current and $j_e^{(1)}$ is the correction to the Ohmic current due to time dependence of the fields. The current $j_H^{(0)}$ is the Hall current in the medium generated due to the perpendicular electric and magnetic fields, and $j_H^{(1)}$ and $j_H^{(2)}$ are the correction to Hall current that comes from the terms ($\dot{\textbf{E}} \times \textbf{B}$) and ($\textbf{E} \times \dot{\textbf{B}}$), respectively.  
\section{Effects of QCD medium interactions and anisotropy}
The hot QCD equation of state (EoS) effect can be incorporated in the analysis through the quasiparticle description of the QCD medium~\cite{PhysRevD.103.014007}. In the present analysis, we employ the EQPM in which the thermal medium interactions are captured by a temperature-dependent fugacity parameter~\cite{Chandra:2011en}.
Considering the fact that large anisotropies arise due to the rapid expansion of the QGP, especially in the initial stages of heavy-ion collisions, the response of the anisotropic medium to the electromagnetic fields needs to be studied. The momentum anisotropy has been seen to have a visible impact on the response of the medium to the constant external fields~\cite{Srivastava:2015via}. We explore the impacts of thermal interaction and momentum anisotropy of the medium to the electric charge transport in the presence of time-varying fields below. 
\begin{figure*}
    \centering
    \centering
    \hspace{-2.5cm}
    \includegraphics[width=0.485\textwidth]{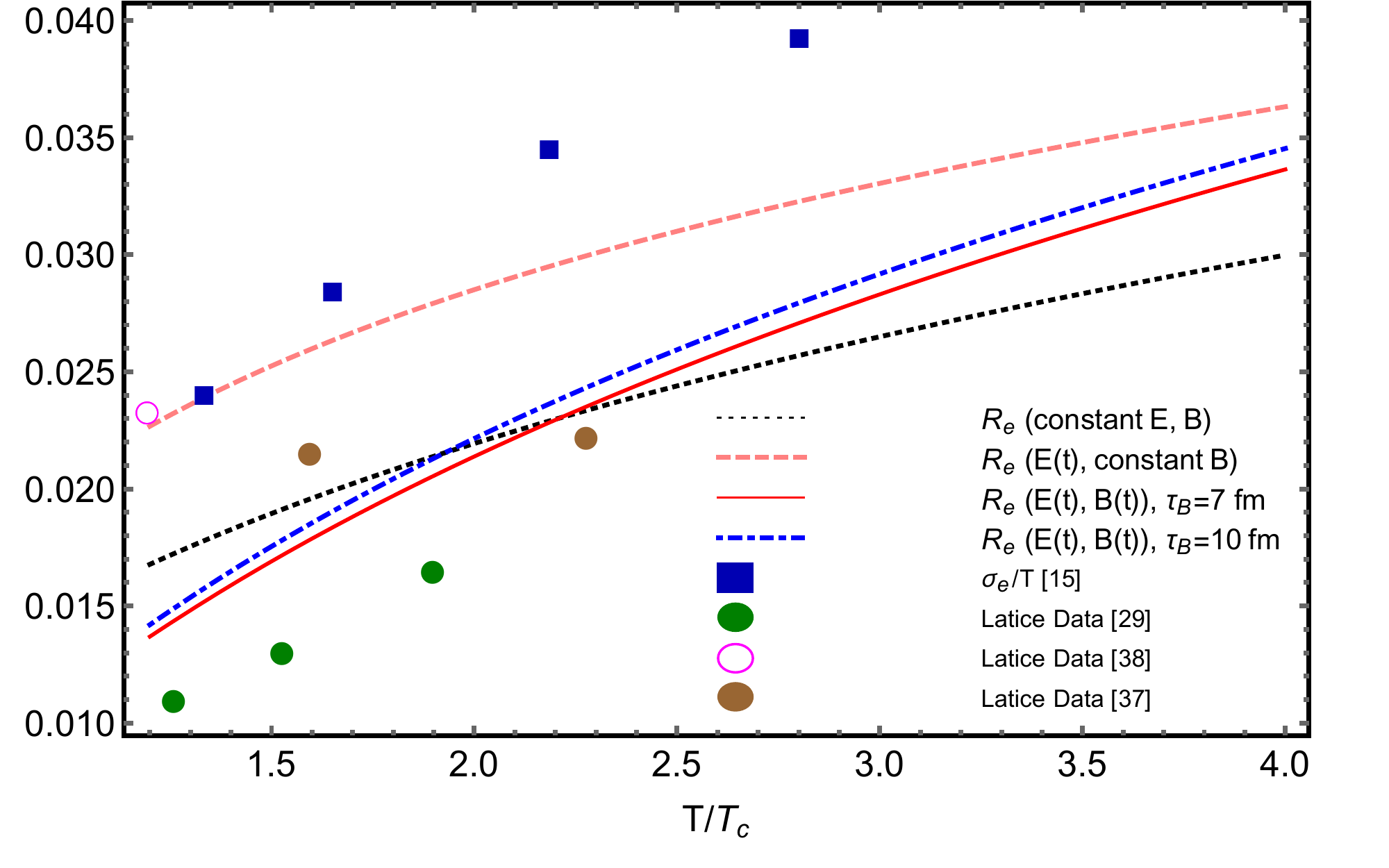}
    \hspace{-.5cm}
    \includegraphics[width=0.571\textwidth]{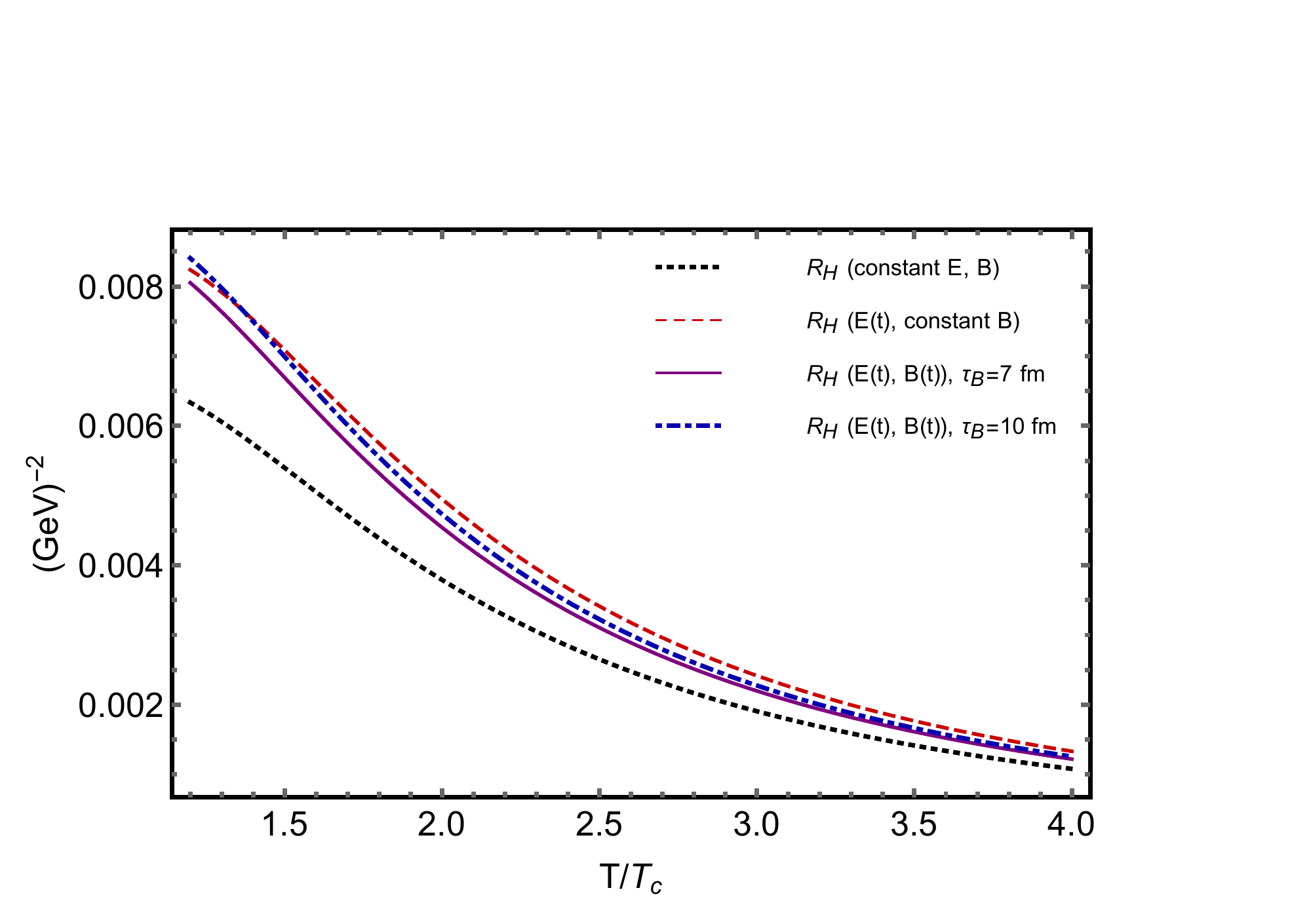}
    \hspace{-2.5cm}
    \caption{\small The temperature dependence of $R_e$ (left panel) and $R_H$ (right panel) for various choices of the external electromagnetic fields: (i). (Constant E, B), (ii).  Time-varying electric and constant magnetic field (E(t), constant B), (iii). Time-varying electromagnetic fields (E(t), B(t))). For constant magnetic field case, $eB = 0.03$ GeV$^2$. The results are compared with lattice data~\cite{Amato:2013naa,Aarts:2007wj,PhysRevD.83.034504} and transport theory estimation~\cite{Thakur:2019bnf} at B=0.}
\label{f1}
\end{figure*}
\subsection{EoS effect: Quasiparticle description}
Within the EQPM prescription, the thermal QCD medium can be described in terms of non-interacting/weakly interacting quasiparticles with the EQPM effective degrees of freedom and modified single-particle energy dispersion as~\cite{Chandra:2011en},
\begin{align}
   &f^0_{k} = \frac{z_{k} e^{\frac{-\epsilon \mp \mu}{T}}}{1 \mp z_{k} e^{\frac{-\epsilon \mp \mu}{T}}}, && \omega_k = \epsilon +\delta \omega_k,
\end{align}
where $z_k$ is the effective fugacity parameter and is related to medium modified part of the energy dispersion as $\delta \omega_k = T^2 \partial_T \ln{z_{k}}$. The near-equilibrium dynamics of the thermal medium can be described within the effective kinetic theory based on the EQPM~\cite{Mitra:2018akk}. The mean-field force term in the effective transport equation, which emerges from in-medium interactions, indeed appears as the mean-field contribution to the transport coefficients associated with the dissipative process. Following the prescriptions of the EQPM kinetic theory, we have the following forms for the components of current density:
\begin{align}
j_{e}^{(0)}=&{E} \sum_k \sum_f (q_{f_k})^2  \int d\Tilde{P}_{1k} \Big\{M_1 -\delta \omega_k \frac{M_1}{p} \Big\},\\
j_{e}^{(1)}=&{-\dot{E}} \sum_k \sum_f (q_{f_k})^2 \int d\Tilde{P}_{1k}  \Big\{M_2 -\delta \omega_k  \frac{M_2}{p} \Big\},\\
j_{H}^{(0)}=&{E}\sum_k \sum_f (q_{f_k})^3  \int d\Tilde{P}_{1k}  \Big\{\frac{M}{\epsilon} -\delta \omega_k \frac{M}{p\epsilon} \Big\},\\
j_{H}^{(1)}=&{-\dot{E}}\sum_k \sum_f (q_{f_k})^3  \int d\Tilde{P}_{1k} \Big\{\frac{M_3}{\epsilon} -\delta \omega_k \frac{M_3}{p\epsilon} \Big\},\\
j_{H}^{(2)}=&\frac{E}{\tau_B}\sum_k \sum_f (q_{f_k})^3  \int d\Tilde{P}_{1k} \tau_R \Big\{\frac{M}{\epsilon} -\delta \omega_k \frac{M}{p\epsilon} \Big\},
\end{align}
where $d\Tilde{P}_{1k} = dP\,\frac{2N_c}{3 \omega_{k}} \frac{p^2}{\epsilon}(-\frac{\partial f^0_k}{\partial \epsilon})$. The term associated with $\delta\omega_k$ represents the mean-field correction term to each component of the current density. Note that at asymptotically high temperature, the medium behaves as ultra-relativistic system with ideal EoS with $z_k\rightarrow 0$ and hence, the mean field contribution vanishes.  
\begin{figure*}
    \centering
    \vspace*{-30mm}
    \includegraphics[width=0.51\textwidth]{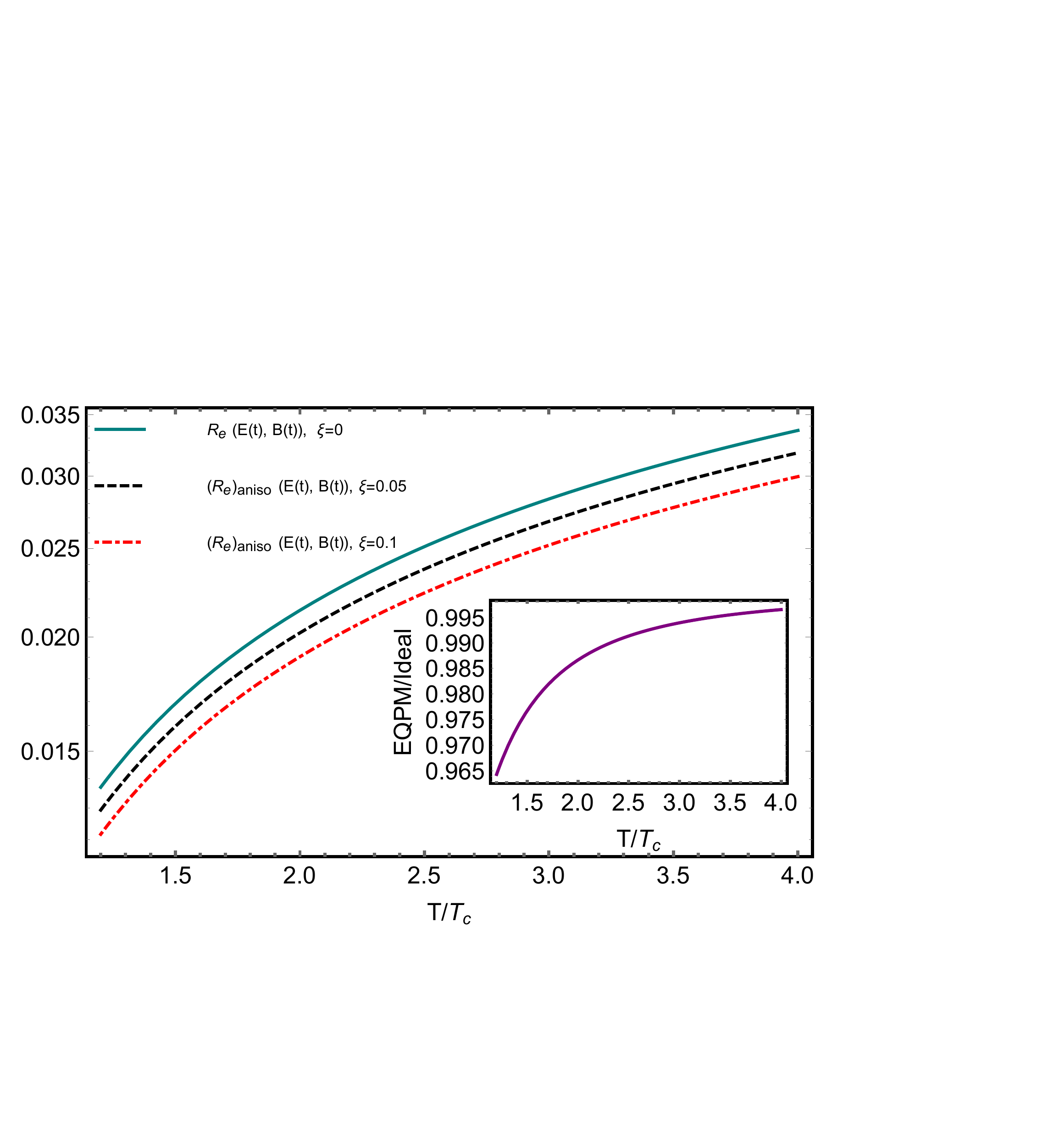}
    \hspace{-1 cm}
    \includegraphics[width=0.53\textwidth]{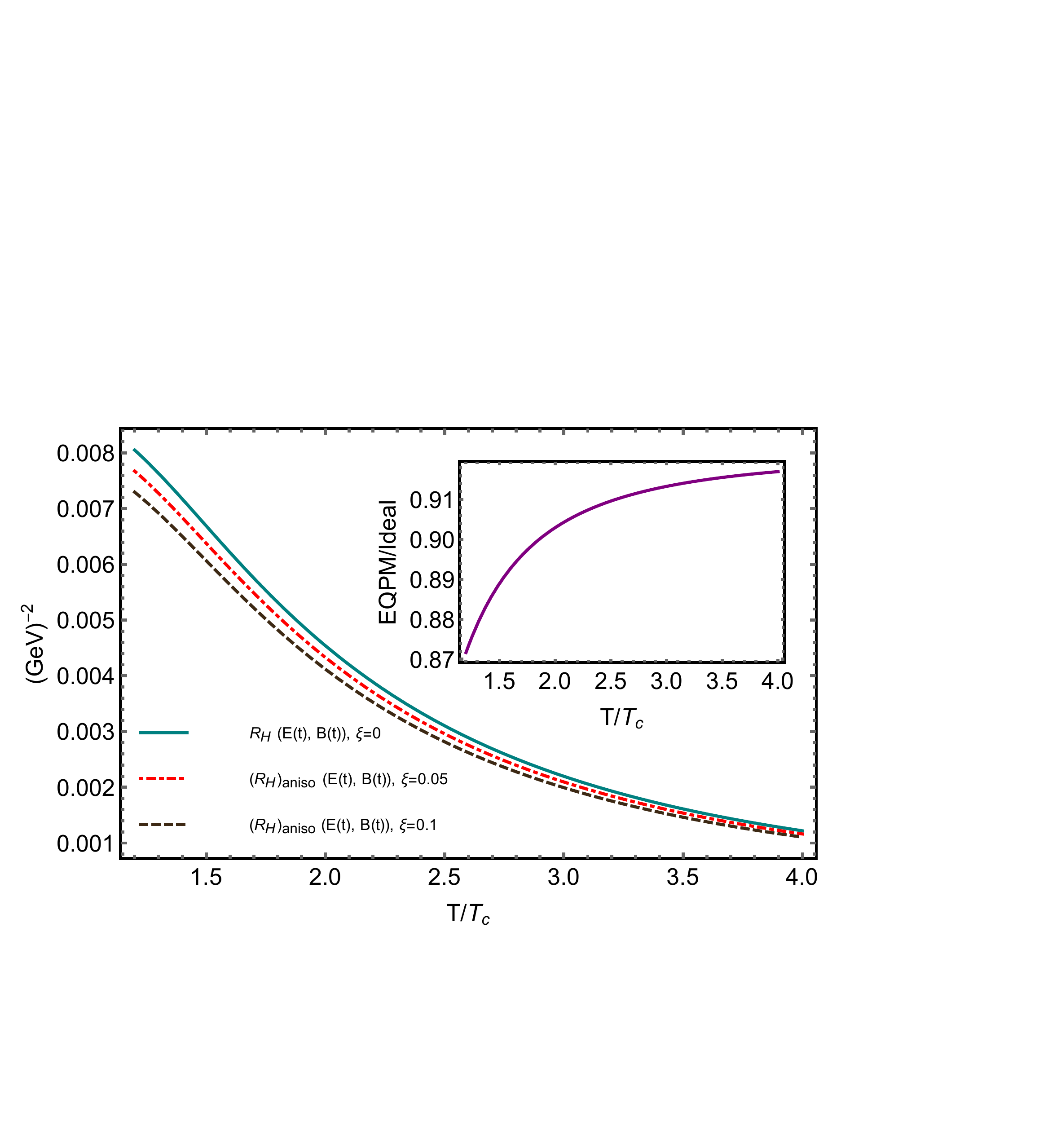}
    \vspace*{-22.5mm}
    \caption{\small Effect of anisotropy and QCD EoS on the temperature behavior of $R_e$ (left panel) and $R_H$ (right panel) in the case of time-varying fields with $\tau_B=7 fm$. $({R_e})_{aniso}$ and $({R_H})_{aniso}$ denote $R_e$ and $R_H$ in an anisotropic medium. }
\label{f2.1}
\end{figure*}
\subsection{Effect of momentum anisotropy}
The physics of anisotropy can be embedded in momentum distribution functions and can be represented in terms of re-scaled isotropic distribution as~\cite{Schenke:2006xu,Romatschke:2003ms},
\begin{align}\label{B.1}
    f_{{(\text{aniso})_k}}= \sqrt{1+\xi}\,f_k^0\Big(\sqrt{p^2+\xi({\bf p}\cdot{\bf n})^2}\Big),
\end{align}
where $\xi$ is the anisotropic parameter and ${\bf n}$ is the direction of anisotropy. For a weakly anisotropic medium, we have $\xi\ll 1$, and Eq.~(\ref{B.1}) reduces to,
\begin{align}
    f_{{(\text{aniso})_k}}= f_k^0-\frac{\xi}{2\epsilon T}({\bf p}\cdot{\bf n})^2 f_k^{0\, 2}e^{\frac{\epsilon \mp \mu}{T}},
\end{align}
with ${\bf p}=(p\sin\theta \cos\phi,\, p\sin\theta \sin\phi,\, p\cos\theta)$ and ${\bf n}=(\cos\alpha,\, 0,\,  \sin\alpha)$. Following the formalism of Ref.~\cite{Srivastava:2015via}, and solving the relativistic Boltzmann equation we obtain the electric current density in the direction of external time-varying electric field in an anisotropic medium as,
\begin{align}
    (j_{e})_{{\text{aniso}}} =j_e^{(0)} +\delta j_e^{(0)} +j_e^{(1)} +\delta j_e^{(1)},
\end{align}
where $j_e^{(0)}$ and $j_e^{(1)}$ are the isotropic components defined in Eq.~(\ref{31}) and Eq.~(\ref{31.1}), respectively. The corrections to the  electric current due to momentum anisotropic of the medium, denoted by $\delta j_e^{(0)}$ and $\delta j_e^{(1)}$, can be defined as,
\begin{align}\label{2.18}
&\delta j_{e}^{(0)}=-\xi\frac{E}{3} N_c\sum_k \sum_f (q_{f_k})^2 l \int_{0}^{\infty} dp L e^{\frac{\epsilon \mp \mu}{T}} M_1,\\
&\delta j_{e}^{(1)}=\xi\frac{\dot{E}}{3}N_c \sum_k \sum_f (q_{f_k})^2 l \int_{0}^{\infty} dp L e^{\frac{\epsilon \mp \mu}{T}} M_2.
\end{align}
with $l = \frac{1}{6\pi^2 T^2} $ and $L = \frac{ p^6}{\epsilon} (f^0_k)^2$. It is important to emphasize that in the absence of magnetic field Eq.~(\ref{2.18}) reduces back to the findings of~\cite{Srivastava:2015via}. Similarly, we define the components of current density in the direction transverse to the fields in the anisotropic medium as,  
\begin{align*}
 (j_{H})_{{\text{aniso}}} =j_H^{(0)} +\delta j_H^{(0)} +j_H^{(1)} +\delta j_H^{(1)}+j_H^{(2)} +\delta j_H^{(2)}.
\end{align*}
The isotropic terms are described in Eqs.~(\ref{31.2})-(\ref{31.4}) and the anisotropic contributions take the following forms,
\begin{align}\label{2.18.3}
&\delta j_{H}^{(0)}=-\xi\frac{E}{3}N_c \sum_k \sum_f (q_{f_k})^3 l \int_{0}^{\infty} dp \frac{L}{\epsilon} e^{\frac{\epsilon \mp \mu}{T}}  M,\\
&\delta j_{H}^{(1)}=\xi\frac{\dot{E}}{3}N_c\sum_k \sum_f (q_{f_k})^3 l \int_{0}^{\infty} dp \frac{L}{\epsilon} e^{\frac{\epsilon \mp \mu}{T}} M_3,\\
&\delta j_{H}^{(2)}=-\xi\frac{E}{3\tau_B}N_c \sum_k \sum_f (q_{f_k})^3 l \int_{0}^{\infty} dp \frac{L\tau_R}{\epsilon} e^{\frac{\epsilon \mp \mu}{T}} M.
\end{align}
\section{Results and Discussions}
We initiate the discussion with the QCD medium response to the time-varying electromagnetic field. The medium response to the fields is quantified in terms of induced current in the direction of the external electric field $j_e$ and in the direction perpendicular to electromagnetic fields $j_H$. We define the following ratios,
\begin{align*}
    &R_{e} =  \frac{j_{e}^{(0)}}{ET} + \frac{j_{e}^{(1)}}{ET},
    &&R_{H} =  \frac{j_{H}^{(0)}}{EBT} + \frac{j_{H}^{(1)}}{EBT} +\frac{j_{H}^{(2)}}{EBT},
\end{align*}
such that ${j_{e}^{(0)}}/({ET)}=\sigma_e/T$ denotes the dimensional less quantity  in the case of constant electromagnetic fields where the term ${j_{e}^{(1)}}/{(ET)}$ gives further correction due to the time dependence of ${\bf E}$. Similarly, ${j_{H}^{(0)}}/({EBT})$ represents the leading order term $\sigma_H/(BT)$ followed by the correction terms with $j_{H}^{(1)}$ and $j_{H}^{(2)}$ in the direction $({\bf\hat{e}}\times{\bf\hat{b}})$. 
The space-time profile of the electric and magnetic fields are described in Ref.~\cite{Hongo:2013cqa} in which the strength of the inhomogenetiy in time can be quantified with decay time  $\tau_{E/B}$ as ${\dot{\textbf{E}}}/{\textbf{E}} \propto \tau_E$ and ${\dot{\textbf{B}}}/{\textbf{B}} \propto \tau_B$. 

The impact of the time dependence of the external fields on the temperature behavior of $R_e$ is depicted in Fig.~\ref{f1} (left panel). We observe that the time dependence of the electric and magnetic fields has a significant impact on the medium response of the system. In the case of a constant magnetic field and time-varying electric field, $R_e$ is higher than that in the case of constant electric and magnetic field due to the extra component associated with ${\bf \dot{E}}$. However, the inclusion of time dependence of the magnetic field introduces back current in the conducting medium. This further affects the temperature dependence of $R_e$. We compare the results with the lattice estimation of $\sigma_e/T$ at vanishing magnetic field and constant electric field as described in  Refs.~\cite{Amato:2013naa,Aarts:2007wj,PhysRevD.83.034504} and also with the transport theory results~\cite{Thakur:2019bnf}. Notably, the effect of inhomogeneity of time of the fields to the current density is seen to be more pronounced in the temperature regime closer to the transition temperature $T_c$.
The effect of additional components to the Hall current due to the time dependence of the fields is studied by demonstrating the temperature dependence of $R_{H}$ in Fig.~\ref{f1} (right panel). Similar to the case of $j_e{\bf \hat{e}}$, the inclusion of time dependence of the electric and magnetic fields has a visible impact on the Hall current, especially in the low-temperature regimes.
    
The impact of the momentum anisotropy on the temperature behavior of $R_e$ and $R_H$ are plotted in Fig.~\ref{f2.1}.
We observe both $j_e{\bf \hat{e}}$ and $j_H({\bf \hat{e}}\times {\bf \hat{b}})$ in the presence of time-varying fields decrease with an increase in anisotropy in the medium. This observation agrees with the study of charge transport with constant electromagnetic fields~\cite{Srivastava:2015via}. We have also explored the effect of in-medium interactions on the charge transport in the QCD medium (inset plots). It is seen that the QCD EoS and the mean-field effects further decrease the current densities. The effect of the EoS is prominent at the low-temperature regimes, especially for the Hall current density. 
\section{Conclusion and Outlook}
In conclusion, we have explored the response of the QCD medium to the time-varying electromagnetic fields. We have obtained a general form of the near-equilibrium distribution function of the medium constituents in the presence of inhomogeneous electromagnetic fields. To that end, we have solved the relativistic Boltzmann equation within the relaxation time approximation. The QCD medium response to the external time dependent fields has been quantified in terms of induced current. Notably, we have obtained additional components to the electrical and Hall current densities that arise from the time dependence of the fields. The impact of the additional terms on the current densities and the respective conductivities are observed to be significant, especially in the temperature regime not very far from $T_c$. 
We have compared the results with the lattice estimations and transport theory results with constant external fields.
Further, we have studied the impacts of the thermal QCD EoS and momentum anisotropy to the electric charge transport in the presence of time-varying fields. The in-medium interaction effects are incorporated in the analysis through the quasiparticle and the followed effective kinetic theory description of the charge transport. It is seen that both the EoS and anisotropic effects to the current densities are non-negligible in the temperature regime near $T_c$.

The additional components to the Ohmic and Hall current densities due to the decay of the electric and magnetic fields in the medium may perhaps play a significant role in the realistic magnetohydrodynamical framework for the QCD medium in the heavy-ion collision experiments. Furthermore, the inclusion of the back current along with the impact of momentum anisotropy of the medium to charge transport within an effective description is essential for the complete understanding of charge-dependent directed flow of final-stage particles in the asymmetric collision experiments~\cite{Gursoy:2018yai}. We intend to explore the phenomenological aspect of the analysis in the near future. The study of thermal and momentum transport in the presence of a time-varying magnetic field in the QCD medium is another interesting direction to explore in the future. 

\section{Acknowledgments}
M.K. would like to acknowledge the Indian Institute of Technology Gandhinagar for the postdoctoral fellowship.


\bibliography{ref}{}

\end{document}